\documentclass[12pt]{article}

\usepackage{graphicx}
\usepackage{amssymb}

\begin{document}

\title{A note on the calculation of the long-wavelength limit of the bosonic excitation spectrum}


\author{Andrij Rovenchak\\
Department for Theoretical Physics,\\ Ivan Franko National University of Lviv,\\
12 Drahomanov St., Lviv 79005, Ukraine \\
e-mail: {andrij.rovenchak@gmail.com}
}

\maketitle

\def\and{;\ }

\begin{abstract}
An approach is proposed to analyze an interacting bosonic system using two-time temperature Green's functions on the collective variables. Two systems are studied: liquid helium-4 and the Yukawa Bose-liquid being a model of the nuclear matter. The suggested decoupling in the equations of motion for Green's functions yields a good description of the elementary excitation spectrum in the long-wavelength limit. 

\textbf{Key words:}
 {Bose-system \and Excitation spectrum \and Liquid helium-4 \and Yukawa Bose-liquid \and Nuclear matter \and Two-time temperature Green's functions}

PACS numbers: {05.30.Jp \and 03.75.Hh \and 67.25.--k \and 21.65.--f}
\end{abstract}

\def\ben{\begin{eqnarray*}}
\def\een{\end{eqnarray*}}
\def\be{\begin{eqnarray}}
\def\ee{\end{eqnarray}}

\def\k{{\bf k}}
\def\q{{\bf q}}
\def\p{{\bf p}}
\def\eps{\varepsilon}
\def\drho{\partial}

\def\GreenF#1#2{\langle\!\langle#1|#2\rangle\!\rangle}

\section{Introduction}
\label{sec:Intro}
In this note, I am going to address a rather classical problem of the low-temperature physics and the theory of Bose-systems specifically. The excitation spectrum of a Bose-liquid was an important element in the formulation of the theory of superfluidity of liquid helium-4. Phenomenologically formulated by Landau \cite{Landau:1941,Landau:1947}, the spectrum was microscopically derived by Bijl \cite{Bijl:1940}, Bogoliubov \cite{Bogoliubov:1947}, Feynman and Cohen \cite{Feynman:1954,Feynman&Cohen:1956}. In the following decades, numerous works on this subject appeared \cite{Jackson&Feenberg:1962,Sunakawa_etal:1969,Vakarchuk&Yukhnovskii:1980,%
Apaja_etal:1998,Kruglov&Collett:2001,Pashitskii_etal:2002,Adamenko_etal:2003,%
Kruglov&Collett:2008,Fak_etal:2012,Bobrov&Trigger:2013}, just to mention a few.

Experimental observation of the Bogoliubov excitations in exciton--polariton Bose-condensates was reported by Utsunomiya \textit{et al.} \cite{Utsunomiya_etal:2008}.
Such systems \cite{Rubo_etal:2006,Deng_etal:2010} constitute another group for probing Bose-condensation, alongside dilute alkali gases \cite{Pethick&Smith:2001}.

Approaches based on the Hamiltonian of an interacting Bose-system by Bogoliubov and Zubarev \cite{Bogoliubov&Zubarev:1955} were used by the present author to calculate the effective mass of the $^4$He atom and the excitation spectrum of liquid helium-4 \cite{Rovenchak:2003FNT,Rovenchak:2005JLTP}. In this work, yet another method is proposed allowing for a proper treatment of the long-wavelength domain
of the excitation spectrum. This approach is tested both for helium-4 with a self-consistently derived interatomic potential \cite{Vakarchuk_etal:2000} and for a Bose-liquid with the Yukawa potential.

The paper is organized as follows. Section~\ref{sec:Green} contains all the required definitions of the Hamiltonian and two-temperature Green's functions, which are further applied in the calculations. Decouplings for Green's functions are suggested in Section~\ref{sec:Decoupling} and general expressions for the excitation spectrum are obtained there. Results of calculations for two systems (liquid helium-4 and the Yukawa Bose-liquid) are given in Section~\ref{sec:Results}. A short discussion in Section~\ref{sec:Disc} concludes the paper.

\section{Green's functions}
\label{sec:Green}
Let us consider the Hamiltonian of a Bose-system of $N$ particles in the volume $V$ interacting via pairwise $\Phi({\bf r}_1-{\bf r}_2)$ and three-particle $\Phi_3({\bf r}_1,{\bf r}_2,{\bf r}_3)$ potentials 
\be\label{eq:H}
H = -\frac{\hbar^2}{2m} \sum_{j=1}^N \Delta_j +
\sum_{1\leq j<l\leq N} \Phi({\bf r}_j-{\bf r}_l)+
\sum_{1\leq j<l<s\leq N} \Phi_3({\bf r}_j,{\bf r}_l,{\bf r}_s),
\ee
where $m$ is the mass of a particle and $\Delta_j$ is the Laplace operator with respect to the $j$th coordinate ${\bf r}_j$. In the case of a Bose-system it is possible to pass to the so-called collective variables
\be\label{eq:rho_drho}
\rho_\k = \frac{1}{\sqrt{N}}\sum_{j=1}^N e^{-i\k{\bf r}_j},
\qquad
\drho_{-\k} = \frac{\partial}{\partial\rho_\k},\qquad
\textrm{where}\quad\k\neq 0.
\ee
Hamiltonian (\ref{eq:H}) becomes as follows \cite{Bogoliubov&Zubarev:1955,Rovenchak:2005CEJP}:
\be\label{eq:Hrho}
H &=& \sum_{\k\neq0}\Big[
\eps_k\left(\rho_\k\drho_{-\k}-\drho_\k\drho_{-\k}\right) +
\frac{n}{2}\tilde\nu_k\rho_\k\rho_{-\k}\Big] \\
&&{}+
\frac{1}{\sqrt{N}}\mathop{\sum_{\k\neq0}\sum_{\q\neq0}}\limits_{\k+\q\neq0}
\Big[
\frac{\hbar^2}{2m}\k\q\,\rho_{\k+\q}\drho_{-\k}\drho_{-\q}+
\frac{n^2}{6}\nu_3(\k,\q)\rho_{\k+\q}\rho_{-\k}\rho_{-\q}
\Big],
\ee
where $\eps_k = \hbar^2k^2/2m$ is the free-particle energy, $n=N/V$ is the particle density, $\nu_k$ and $\nu_3(\k,\q)$ are the Fourier transforms of the pairwise and three-particle potentials, respectively, and
\be
\tilde\nu_k = \nu_k +n\nu_3(\k,-\k) -\frac{1}{V}\sum_{\q\neq0}\nu_3(\k,\q).
\ee

With operators $A$ and $B$ written in the Heisenberg representation, the two-time temperature Green's functions are defined as \cite{Zubarev:1974}
\be
\GreenF{A(t)}{B(t')} = i\theta(t-t')\left\langle[A(t),B(t')]\right\rangle,
\ee
where $\theta(t)$ is the Heaviside step function and $[\cdot,\cdot]$ denotes the commutator. 

In the case of $t=t'$, the equations of motion in the frequency representation are given by
\be
\hbar\omega\GreenF{A}{B} = \frac{1}{2\pi}\langle[A,B]\rangle+
\GreenF{[A,H]}{B},
\ee
which upon simple transformations with Eq.~(\ref{eq:Hrho}) leads to the following set:
\be
&&\hbar\omega\GreenF{\rho_\k}{\rho_{-\k}} = 
-\eps_k\GreenF{\rho_\k}{\rho_{-\k}}
+2\eps_k\GreenF{\drho_\k}{\rho_{-\k}} \nonumber\\
&&\qquad\qquad\qquad\qquad{}-\frac{2}{\sqrt{N}} \mathop{\sum_{\q\neq0}}\limits_{\k+\q\neq0}
\frac{\hbar^2}{2m}\k\q\GreenF{\rho_{\k+\q}\drho_{-\q}}{\rho_{-\k}}, \nonumber\\ [-6pt]
\label{eq:hwGreen}\\[-6pt]
&&\hbar\omega\GreenF{\drho_\k}{\rho_{-\k}} = \frac{1}{2\pi}
+\eps_k\GreenF{\drho_\k}{\rho_{-\k}}
+n\nu_k\GreenF{\rho_\k}{\rho_{-\k}} \nonumber\\
&&\quad{}+\frac{1}{\sqrt{N}} \mathop{\sum_{\q\neq0}}\limits_{\k+\q\neq0}
\Bigg[
\frac{\hbar^2}{2m}\k\q\GreenF{\drho_{\k+\q}\drho_{-\q}}{\rho_{-\k}}
-\frac{n^2}{2}\nu_3(\k,\q)\GreenF{\rho_{\k+\q}\rho_{-\q}}{\rho_{-\k}}
\Bigg]
. \nonumber
\ee

Dropping off three-operator Green's functions, i.\,e. in the random phase approximation (RPA), one obtains the following expression for one of the solutions of set (\ref{eq:hwGreen}):
\be
\GreenF{\rho_\k}{\rho_{-\k}} = \frac{\eps_k}{\pi}
\frac{1}{(\hbar\omega)^2-\eps_k^2\alpha_k^2}
\ee
immediately yielding the Bogoliubov spectrum from the poles of Green's functions with respect to $\hbar\omega$:
\be\label{eq:EkRPA}
E_k = \eps_k\alpha_k,
\qquad\textrm{where}\quad
\alpha_k=\left(1+2n\nu_k/\eps_k\right)^{1/2}.
\ee 

While it is also possible to derive the equations of motion for three-operator functions as well, no closed-form expression for the excitation spectrum can be obtained with them. So, another option leading to easier calculations of the spectrum is considered in the next Section.

\section{Green's function decoupling and excitation spectrum}
\label{sec:Decoupling}
Since Green's functions of the $\GreenF{AB}{C}$ type are used to calculate averages of triple products $\langle CAB \rangle$, the following decoupling can be used
\be
\GreenF{A_1B_2}{C_3} = 
\lambda(1,2)\GreenF{A_{1+2}}{C_3}+\mu(1,2)\GreenF{B_{1+2}}{C_3}
\ee
with
\be\label{eq:lambda_mu}
\lambda(1,2) = 
(1-\eta) \frac{\langle C_3A_1B_2\rangle}{\langle C_3A_{1+2}\rangle},
\qquad
\mu(1,2) = \eta  \frac{\langle C_3A_1B_2\rangle}{\langle C_3B_{1+2}\rangle},
\ee
where the value of the parameter $\eta=0\div1$ will be fixed on a later stage.

The following two approximation for averages were tested
\be
\langle C_3A_1B_2\rangle_s = \frac16\bigg(
\langle CA\rangle_1\langle CB\rangle_2\langle AB\rangle_3+
\textrm{symmetrization over indices}
\bigg)
\ee
with six items in the parentheses, hence the ``$s$'' index, and
\be
\langle C_3A_1B_2\rangle_f &=& \frac14\bigg[
\Big(\langle AB\rangle_1\langle CA\rangle_2+
\langle AB\rangle_2\langle CA\rangle_1\Big)\langle CB\rangle_3\\
&&\ {}+\Big(\langle AB\rangle_1\langle CB\rangle_2+
\langle AB\rangle_2\langle CB\rangle_1\Big)\langle CA\rangle_3
\bigg]\nonumber
\ee
with four items in the parentheses.

To decouple the functions entering Eqs.~(\ref{eq:hwGreen}) the following averages are required:
\be
\langle\rho_{-\k}\rho_{\k+\q}\rho_{-\q}\rangle_{s,f} = \frac{1}{\sqrt{N}}
\langle\rho_{-\q}\rho_{\q}\rangle
\langle\rho_{-\k-\q}\rho_{\k+\q}\rangle
\langle\rho_{-\k}\rho_{\k}\rangle,
\ee
which corresponds to the so-called convolution approximation and is identical for both the suggested decoupling types,
\be
&&\langle\rho_{-\k}\rho_{\k+\q}\drho_{-\q}\rangle_{s} = \frac{1}{3\sqrt{N}}
\Big(
  \langle\rho_{-\q}\rho_{\q}\rangle
  \langle\rho_{-\k-\q}\drho_{\k+\q}\rangle
  \langle\rho_{-\k}\drho_{\k}\rangle
\\
&&\quad{}+ 
  \langle\rho_{-\q}\drho_{\q}\rangle
  \langle\rho_{-\k-\q}\rho_{\k+\q}\rangle
  \langle\rho_{-\k}\drho_{\k}\rangle
 +
  \langle\rho_{-\q}\drho_{\q}\rangle
  \langle\rho_{-\k-\q}\drho_{\k+\q}\rangle
  \langle\rho_{-\k}\rho_{\k}\rangle
\Big),\nonumber\\[12pt] 
&&\langle\rho_{-\k}\drho_{\k+\q}\drho_{-\q}\rangle_{s} = \frac{1}{3\sqrt{N}}
\Big(
  \langle\drho_{-\q}\drho_{\q}\rangle
  \langle\rho_{-\k-\q}\drho_{\k+\q}\rangle
  \langle\rho_{-\k}\drho_{\k}\rangle
\\
&&\quad{}+ 
  \langle\rho_{-\q}\drho_{\q}\rangle
  \langle\drho_{-\k-\q}\drho_{\k+\q}\rangle
  \langle\rho_{-\k}\drho_{\k}\rangle
 +
  \langle\rho_{-\q}\drho_{\q}\rangle
  \langle\rho_{-\k-\q}\drho_{\k+\q}\rangle
  \langle\drho_{-\k}\drho_{\k}\rangle
\Big)\nonumber
\ee
and
\be
&&\langle\rho_{-\k}\rho_{\k+\q}\drho_{-\q}\rangle_{f}  \\
&&\quad= 
\frac{1}{2\sqrt{N}}
\Big(
  \langle\rho_{-\q}\rho_{\q}\rangle
  \langle\rho_{-\k-\q}\drho_{\k+\q}\rangle
  \langle\rho_{-\k}\drho_{\k}\rangle
+
  \langle\rho_{-\q}\drho_{\q}\rangle
  \langle\rho_{-\k-\q}\rho_{\k+\q}\rangle
  \langle\rho_{-\k}\drho_{\k}\rangle
\Big),\nonumber\\[12pt] 
&&\langle\rho_{-\k}\drho_{\k+\q}\drho_{-\q}\rangle_{f} \\
&&\quad= 
\frac{1}{2\sqrt{N}}
\Big(
  \langle\drho_{-\q}\drho_{\q}\rangle
  \langle\rho_{-\k-\q}\drho_{\k+\q}\rangle
  \langle\rho_{-\k}\drho_{\k}\rangle
+
  \langle\rho_{-\q}\drho_{\q}\rangle
  \langle\drho_{-\k-\q}\drho_{\k+\q}\rangle
  \langle\rho_{-\k}\drho_{\k}\rangle
\Big).
\nonumber
\ee

Expressions for pairwise averages are easily obtained from the solutions of Eqs.~(\ref{eq:hwGreen}) in RPA as follows \cite{Rovenchak:2003FNT}:
\be
&&\langle\rho_{-\k}\rho_{\k}\rangle = \frac{1}{\alpha_k}, \qquad
\langle\rho_{-\k}\drho_{\k}\rangle = \frac12\left(\frac{1}{\alpha_k}-1\right),\nonumber\\
&&\langle\drho_{-\k}\drho_{\k}\rangle = 
\frac14\left(\frac{1}{\alpha_k}-\alpha_k\right).
\ee

With the decouplings applied, equations of motion (\ref{eq:hwGreen}) become:
\be
\hbar\omega\GreenF{\rho_\k}{\rho_{-\k}} &=& 
-\eps_k^{(1)}\GreenF{\rho_\k}{\rho_{-\k}}
+2\eps_k^{(2)}\GreenF{\drho_\k}{\rho_{-\k}} \nonumber\\
\label{eq:hwGreen_decoupled}\\[-6pt]
\hbar\omega\GreenF{\drho_\k}{\rho_{-\k}} &=& \frac{1}{2\pi}+
\eps_k^{(3)}\GreenF{\drho_\k}{\rho_{-\k}}
+n\nu_k^{(*)}\GreenF{\rho_\k}{\rho_{-\k}}, \nonumber
\ee
where
\be\label{eq:eps1}
&&\eps_k^{(1)} = \eps_k+\frac{2}{\sqrt{N}} \mathop{\sum_{\q\neq0}}\limits_{\k+\q\neq0}
  \frac{\hbar^2}{2m}\k\q\,X(\k,\q),\\
\label{eq:eps2}
&&\eps_k^{(2)} = \eps_k+\frac{1}{\sqrt{N}} \mathop{\sum_{\q\neq0}}\limits_{\k+\q\neq0}
  \frac{\hbar^2}{2m}\k\q\,Y(\k,\q),
\\
\label{eq:eps3}
&&\eps_k^{(3)} = \eps_k + \frac{2}{\sqrt{N}} \mathop{\sum_{\q\neq0}}\limits_{\k+\q\neq0}
  \frac{\hbar^2}{2m}\k\q\,Z(\k,\q),
\\
&&\nu_k^{(*)} = \tilde\nu_k+\frac{1}{\sqrt{N}}
  \mathop{\sum_{\q\neq0}}\limits_{\k+\q\neq0}\nu_3(\k,\q) T(\k,\q).
\ee
The notations used in the above definitions are as follows:
\be
&&X(\k,\q) = (1-\eta)\frac{\langle\rho_{-\k}\rho_{\k+\q}\drho_{-\q}\rangle}{\langle\rho_{-\k}\rho_{\k}\rangle}, \nonumber
\qquad
Y(\k,\q) = \eta \frac{\langle\rho_{-\k}\rho_{\k+\q}\drho_{-\q}\rangle}{\langle\rho_{-\k}\drho_{\k}\rangle}, \nonumber\\[-6pt]
\\
&&Z(\k,\q) = \frac{\langle\rho_{-\k}\drho_{\k+\q}\drho_{-\q}\rangle}{\langle\rho_{-\k}\drho_{\k}\rangle},
\qquad
T(\k,\q) = \frac{\langle\rho_{-\k}\rho_{\k+\q}\rho_{-\q}\rangle}{\langle\rho_{-\k}\rho_{\k}\rangle}\nonumber
\ee

The excitation spectrum is thus given by:
\be\label{eq:EkG3}
E_k = \frac12\left\{\eps_k^{(3)}-\eps_k^{(1)} + 
\sqrt{\left[\eps_k^{(1)}\right]^2+\left[\eps_k^{(3)}\right]^2
+2\eps_k^{(1)}\eps_k^{(3)}+8\eps_k^{(2)}n\nu_k^{(*)}}\right\}
\ee
With the items containing the summations dropped, Bogoliubov's result (\ref{eq:EkRPA}) immediately follows from this expression.

\section{Results}\label{sec:Results}
The calculations of the excitation spectrum were made for two bosonic systems. The first one is the liquid helium-4 and the second one is the Yukawa Bose-liquid with parameters corresponding to the nuclear matter. As data about the details of three-particle interactions in these systems are rather scarce, the contributions of $\nu_3(\k,\q)$ are neglected. It was estimated in \cite{Rovenchak:2005JLTP} that in case of helium-4 such an approach does not influence the long-wavelength limit of the excitation spectrum significantly.

To facilitate the numerical analysis, summation in (\ref{eq:eps1})--(\ref{eq:eps3}) is substituted with integration in the wave-vector space according to such a rule:
\be
\frac{1}{N}\sum_{\q}\ldots = \frac{1}{n}\int d\q\,\ldots.
\ee

The following set of parameters is used for the calculations of the helium-4 excitation spectrum:
\be
m = 4.0026\ \textrm{a.\,m.\,u.},\qquad
n = 0.02185\ \textrm{\AA}^{-3}.
\ee
The data for the interatomic potential $\nu_k$ are taken from \cite{Vakarchuk_etal:2000}.

\begin{figure}[h]
\includegraphics[width=0.48\textwidth]{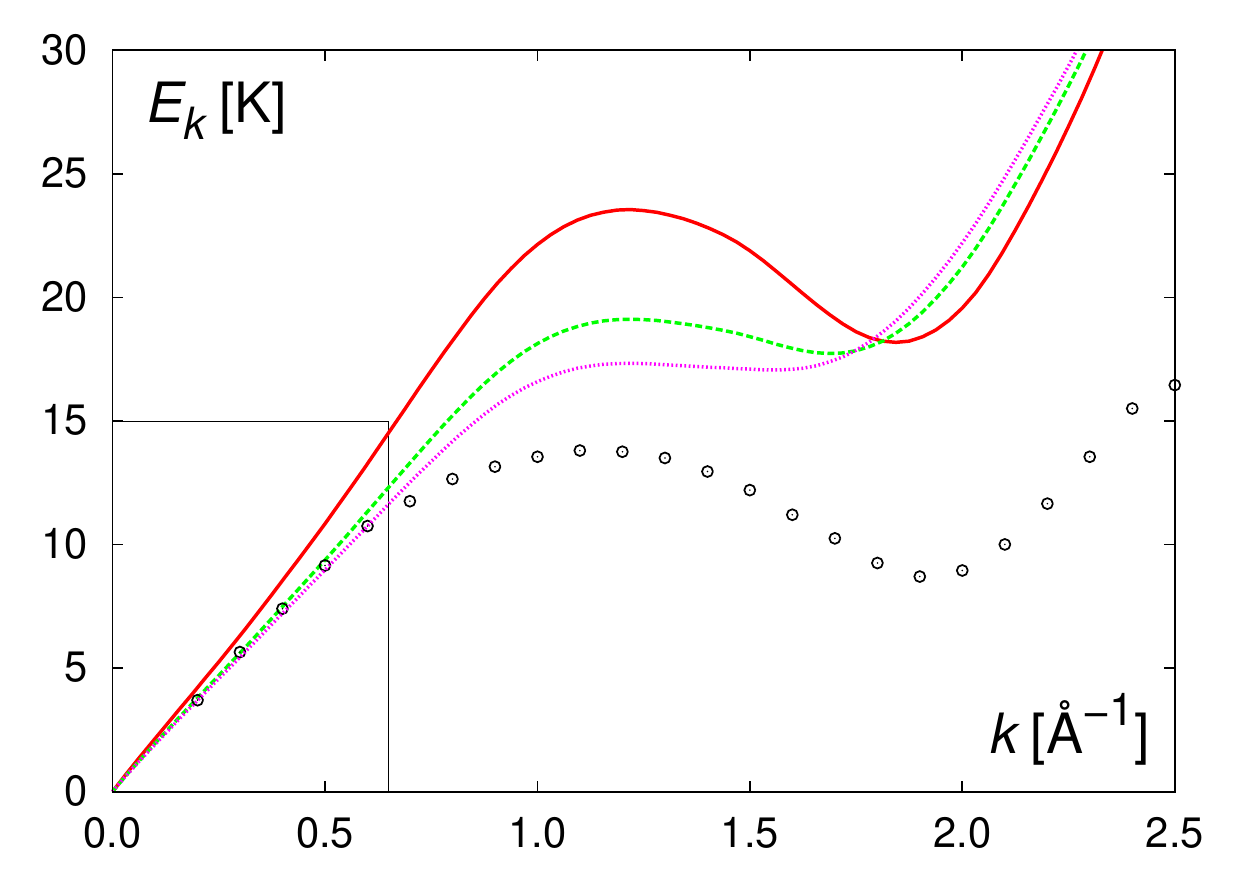}
\quad
\includegraphics[width=0.48\textwidth]{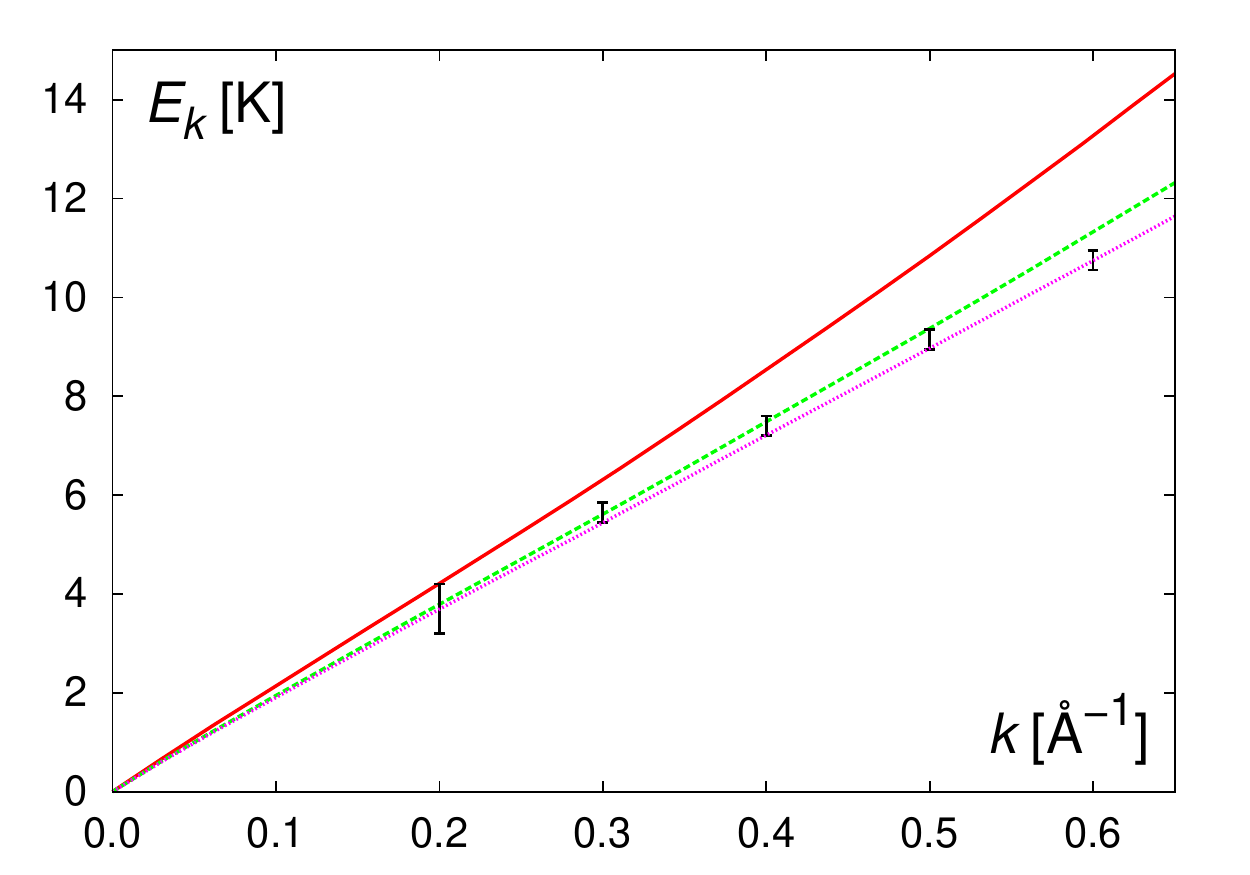}
\caption{(Color online.) Excitation spectrum of the liquid helium-4. The right graph is the enlarged view of the marked rectangular area of the left graph. 
Red solid line -- RPA result (Bogoliubov's spectrum);
green dashed line -- spectrum (\ref{eq:EkG3}) with the $s$-type decoupling, $\eta = 1$;
magenta dotted line -- spectrum (\ref{eq:EkG3}) with the $f$-type decoupling, $\eta = 1$.
Circles (errorbars) are the experimental data from \cite{Cowley&Woods:1971}.}
\label{fig:EkHe}
\end{figure}

Results for the excitation spectrum of helium-4 according to Eq.~(\ref{eq:EkG3}) are shown in Fig.~\ref{fig:EkHe} compared to the RPA approximation (Bogoliubov's spectrum) and experimental data. The $\eta$ parameter is set $\eta=1$ as for smaller values the correction to the RPA result appear insufficient to produce the correct slope of the phonon (linear at $k\to 0$) branch. Both of the suggested decoupling types are found suitable to describe the long-wavelength behavior of the helium-4 spectrum without introduction of the effective mass, which is required in the RPA, cf. \cite{Rovenchak:2003FNT}. On the other hand, the proposed method still fails at higher values of the wave vector and further modifications should be sought for to reproduce the maxon and roton domains successfully in this approach.

Another interesting problem for analysis is the nuclear matter, where different types of Bose-condensation are predicted \cite{Cleymans&vonOertzen:1990,Rezaeian&Pirner:2006}. It is possible to model the nuclear matter as a Bose-liquid interacting via the Yukawa potential $\Phi(r) = \epsilon\,e^{-r/\sigma}/r$ \cite{Ceperley&Chester:1976}. The Fourier transform of this potential reads \cite{Strepparola_etal:1998}
\be
\nu_k = \frac{4\pi\epsilon\sigma^3}{1+\sigma^2k^2}.
\ee
with the following values of the parameters \cite{Ceperley&Chester:1976}:
\be
\epsilon = 5725\ \textrm{MeV},\qquad
\sigma = 0.244\ \textrm{fm}^{-1}.
\ee
Other quantities used to model the nuclear matter are as follows:
\be
\Lambda^* = \frac{2\pi\hbar}{\sigma\sqrt{\epsilon m}} = 1.08,\qquad
n = 0.16\ \textrm{fm}^{-3}. 
\ee

\begin{figure}[h]
\includegraphics[width=0.48\textwidth]{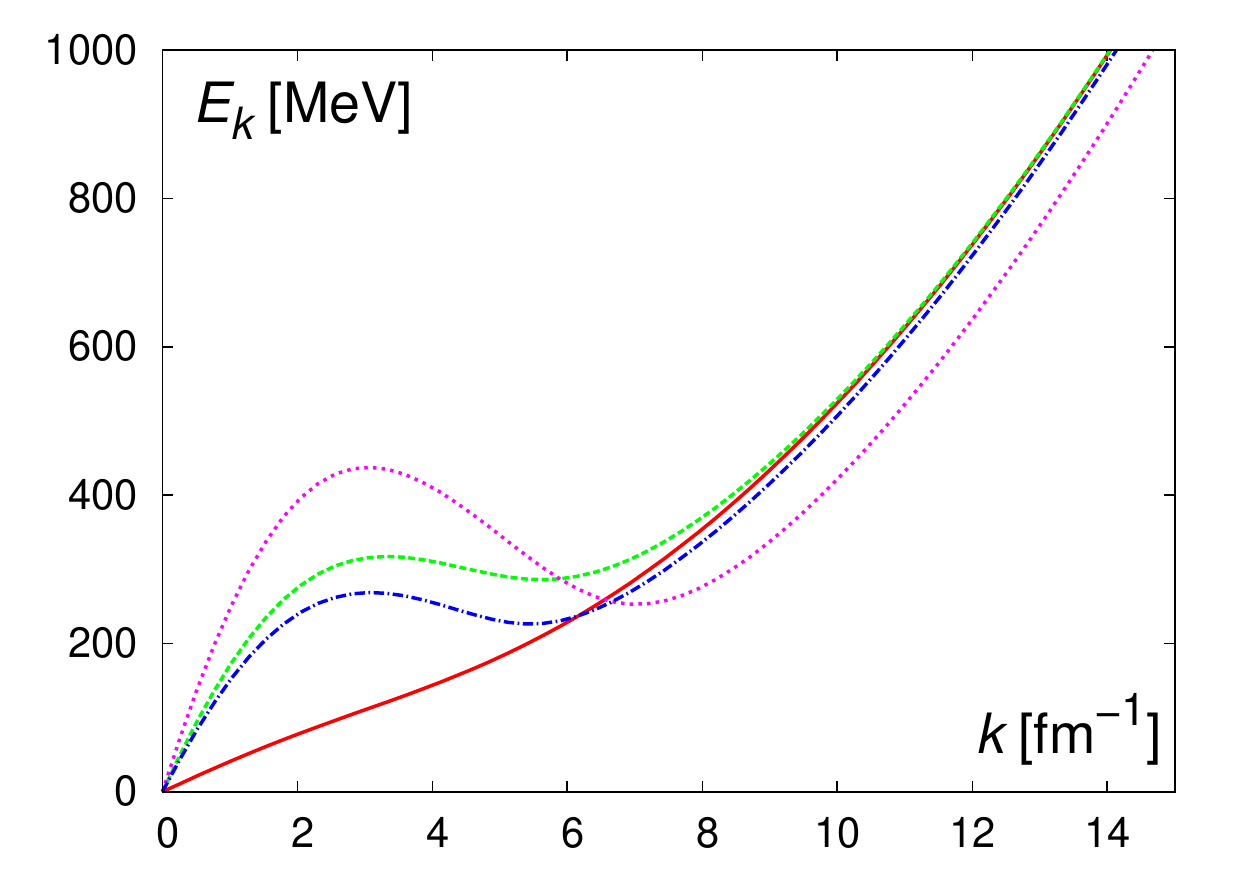}
\caption{(Color online.) Excitation spectrum of the Yukawa Bose liquid with parameters corresponding to the nuclear matter.
Red solid line -- RPA result (Bogoliubov's spectrum);
green dashed line -- spectrum (\ref{eq:EkG3}) with the $s$-type decoupling, $\eta = 0$;
magenta dotted line -- spectrum (\ref{eq:EkG3}) with the $f$-type decoupling, $\eta = 0$;
blue dashed-dotted line -- spectrum (\ref{eq:EkG3}) with the $s$-type decoupling, $\eta = 0.1$
}
\label{fig:EkYukawa}
\end{figure}

Results for the excitation spectrum are given in Fig.~\ref{fig:EkYukawa}. The obtained correction to the RPA changes the shape of the $E_k$ curve leading to a good qualitative agreement with other data \cite{Strepparola_etal:1998,Halinen_etal:2000}. It should be mentioned that the values of the $\eta$ parameter are close to zero in this case as for $\eta\gtrsim0.2$ unphysical divergences in the domain of the minimum ($k=4\div6$~fm$^{-1}$) appear.

\section{Discussion}\label{sec:Disc}
In summary, an approach was proposed to treat an interacting bosonic system using two-time temperature Green's functions on the collective variables leading to a good description of the elementary excitation spectrum in the long-wavelength limit. For two models considered in the work different values of the Green's function decoupling parameter $\eta$ should be taken: $\eta=1$ for the liquid helium-4 and $\eta=0$ for the Yukawa Bose-liquid.

General expression obtained in the work can be further applied to study other bosonic systems, especially where the Fourier transforms of the interatomic potential are known. This includes dilute Bose-gases, where the $\delta$-function potential is applicable, cf. similar analysis \cite{Vorobets:2013} for a binary bosonic mixture, and charged Bose-systems \cite{Panda&Panda:2010}.


\end{document}